\documentclass[reprint, pra,
amsmath,amssymb,showpacs,superscriptaddress]{revtex4-1}
\pdfoutput=1

\usepackage{siunitx}
\usepackage[breaklinks=false]{hyperref}
\usepackage{graphicx}
\usepackage{braket}
\usepackage{verbatim}
\usepackage{float}
\usepackage{amsmath}

\begin{document}
\title{Controlling the Rotational and Hyperfine State of Ultracold $^{87}$Rb$^{133}$Cs Molecules}

\author{Philip D. Gregory}
\affiliation{Joint Quantum Centre (JQC) Durham-Newcastle, Department of
Physics, Durham University, South Road, Durham DH1 3LE, United Kingdom}

\author{Jesus Aldegunde}
\affiliation{Departamento de Quimica Fisica, Universidad de Salamanca, 37008
Salamanca, Spain}

\author{Jeremy M. Hutson}
\email{J.M.Hutson@durham.ac.uk} \affiliation{Joint Quantum Centre (JQC)
Durham-Newcastle, Department of Chemistry, Durham University, South Road,
Durham, DH1 3LE, United Kingdom}

\author{Simon L. Cornish}
\email{S.L.Cornish@durham.ac.uk} \affiliation{Joint Quantum Centre (JQC)
Durham-Newcastle, Department of Physics, Durham University, South Road, Durham
DH1 3LE, United Kingdom}

\begin{abstract}

We demonstrate coherent control of both the rotational and hyperfine state of
ultracold, chemically stable $^{87}$Rb$^{133}$Cs molecules with external
microwave fields. We create a sample of $\sim2000$ molecules in the lowest
hyperfine level of the rovibronic ground state $N=0$. We measure the transition
frequencies to 8 different hyperfine levels of the $N=1$ state at two magnetic
fields $\sim23$~G apart. We determine accurate values of rotational and
hyperfine coupling constants that agree well with previous calculations. We
observe Rabi oscillations on each transition, allowing complete population
transfer to a selected hyperfine level of $N=1$. Subsequent application of a
second microwave pulse allows transfer of molecules back to a different
hyperfine level of $N=0$.

\end{abstract}

\maketitle

%Catchy introductory paragraph about molecules%
Ultracold heteronuclear molecules can provide many exciting new avenues of
research in the fields of quantum state controlled
chemistry~\cite{Ospelkaus:2010b, Krems:2008}, quantum
information~\cite{DeMille:2002}, quantum simulation~\cite{Santos:2002,
Baranov:2012}, and precision measurement~\cite{Flambaum:2007, Isaev:2010,
Hudson:2011}. The large electric dipole moments accessible in such systems
allow interactions to be tuned over length scales similar to the spacing
between sites in an optical lattice. As such, this is an area of intense
research with multiple groups recently reporting the production of dipolar
molecules at ultracold temperatures~\cite{Ni:2008, Takekoshi:2014, Molony:2014,
Park:2015, Guo:2016}.

%Why is state control important? What has been done in the past?%
Full control of the quantum state has been an invaluable tool in ultracold atom
physics; it is therefore highly important to develop similar methods for
ultracold molecules which address the complex rotational and hyperfine
structure. Such control is at the heart of nearly all proposals for
applications of ultracold polar molecules. For example, the rotational states
of molecules might be used as pseudo-spins to simulate quantum
magnetism~\cite{Barnett:2006, Gorshkov:2011}. This requires a coherent
superposition of opposite-parity states to generate dipolar
interactions~\cite{Barnett:2006}, which may be probed by microwave
spectroscopy~\cite{Hazzard:2011, Yan:2013}. Similarly, hyperfine states in the
rotational ground state have been proposed as potential qubits for quantum
computation~\cite{DeMille:2002, Andre:2006, Park:2016}. In this context, robust
coherent transfer between the hyperfine states is essential. Such transfer can
be achieved using a scheme proposed by Aldegunde~\emph{et al.}\
\cite{Aldegunde:2009} which employs microwave fields to manipulate the
molecular hyperfine states. This approach has been implemented for the
fermionic heteronuclear molecules $^{40}$K$^{87}$Rb~\cite{Ospelkaus:2010,
Neyenhuis:2012} and $^{23}$Na$^{40}$K~\cite{Will:2016}, leading to
ground-breaking studies of the dipolar spin-exchange
interaction~\cite{Yan:2013} and nuclear spin coherence time~\cite{Park:2016}.

%Outline of the paper%
In this letter, we report microwave spectroscopy of bosonic $^{87}$Rb$^{133}$Cs
in its ground vibrational state, and coherent state transfer from the absolute
rovibrational and hyperfine ground state to a chosen well-defined single
hyperfine state in either the first-excited or ground rotational states. We
demonstrate the high precision with which we can map out the rotational energy
structure of the $^{87}$Rb$^{133}$Cs molecule in the lowest vibrational state.
We use our measurements to obtain new values for the rotational constant,
scalar spin-spin coupling constant, electric quadrupole coupling constants, and
nuclear $g$-factors (including shielding) for the molecule. Microwave
$\pi$-pulses are used to transfer the molecules first to a single hyperfine
level of the first-excited rotational state, then back to a different hyperfine
level of the rovibrational ground state.

%Description of the Hamiltonian used to describe the system
We calculate the energy level structure of $^{87}$Rb$^{133}$Cs in the
electronic and vibrational ground state by diagonalizing the relevant
Hamiltonian. In the presence of an externally applied magnetic field, this
Hamiltonian ($H$) can be decomposed into rotational~($H_{\text{r}}$),
hyperfine~($H_{\text{hf}}$), and Zeeman~($H_{\text{Z}}$)
components~\cite{Ramsay:1952, Brown:2003, Bryce:2003, Aldegunde:2008}
\begin{equation}
H=H_{\text{r}}+H_{\text{hf}}+H_{\text{Z}}.
\label{eqn:Hamiltonian}
\end{equation}
These contributions are given by
\begin{subequations}
\begin{align}
H_{\text{r}}&=B_{v}\boldsymbol{N}^{2}-D_{v}\boldsymbol{N}^{2}\boldsymbol{N}^{2}, \label{eqn:Rotational} \\
H_{\text{hf}}&= \sum_{i=\text{Rb}, \text{Cs}}{\boldsymbol{V}_{i}\cdot\boldsymbol{Q}_{i}}
+ \sum_{i=\text{Rb}, \text{Cs}}{c_{i}\boldsymbol{N}\cdot\boldsymbol{I}_{i}} \nonumber \\
 & ~~~~~~~~~~~~~~~~+ c_{3}\boldsymbol{I}_{\text{Rb}}\cdot\boldsymbol{T}\cdot\boldsymbol{I}_{\text{Cs}}
 + c_{4}\boldsymbol{I}_{\text{Rb}}\cdot\boldsymbol{I}_{\text{Cs}}, \label{eqn:Hyperfine}\\
H_{\text{Z}}&= -g_{r}\mu_{\rm N}\boldsymbol{N}\cdot\boldsymbol{B}
- \sum_{i=\text{Rb}, \text{Cs}}{g_{i}(1-\sigma_{i})\mu_{\rm N}\boldsymbol{I}_{i}\cdot\boldsymbol{B}}. \label{eqn:Zeeman}
\end{align}
\end{subequations}
The rotational contribution (Eqn.~\ref{eqn:Rotational}) is defined by the
rotational angular momentum of the molecule $\boldsymbol{N}$, and the
rotational and centrifugal distortion constants $B_{v}$ and $D_{v}$. The
hyperfine contribution (Eqn.~\ref{eqn:Hyperfine}) consists of four terms. The
first term describes the electric quadrupole interaction with coupling
constants $(eqQ)_{\text{Rb}}$ and $(eqQ)_{\text{Cs}}$. The second term is the
interaction between the nuclear magnetic moments and the magnetic field
generated by the rotation of the molecule, with spin-rotation coupling
constants $c_{\text{Rb}}$ and $c_{\text{Cs}}$. The final two terms represent
the tensor and scalar interactions between the nuclear magnetic moments, with
tensor and scalar spin-spin coupling constants $c_{3}$ and $c_{4}$
respectively. Finally, the Zeeman contribution (Eqn.~\ref{eqn:Zeeman}) consists
of two terms which represent the rotational and nuclear interaction with an
externally applied magnetic field. The rotation of the molecule produces a
magnetic moment which is characterized by the rotational $g$-factor of the
molecule ($g_{r}$). The nuclear interaction similarly depends on the nuclear
$g$-factors ($g_{\text{Rb}}$,~$g_{\text{Cs}}$) and nuclear shielding
($\sigma_{\text{Rb}}$,~$\sigma_{\text{Cs}}$) for each species. We do not apply
electric fields in this work, which would require the addition of a further
Stark contribution to the Hamiltonian and significantly complicate the spectra
\cite{Ran:2010}.

%Discussion of good quantum numbers
The nuclear spins in $^{87}$Rb$^{133}$Cs are $I_{\text{Rb}}=3/2$
and $I_{\text{Cs}}=7/2$. At zero field, the total angular momentum
$\boldsymbol{F}=\boldsymbol{N}+\boldsymbol{I}_{\text{Rb}} +
\boldsymbol{I}_{\text{Cs}}$ is conserved. For the rotational ground state
($N=0$), the total nuclear spin $\boldsymbol{I} = \boldsymbol{I}_{\text{Rb}} +
\boldsymbol{I}_{\text{Cs}}$ is very nearly conserved, and there are 4 hyperfine
states with $I=2$, 3, 4 and 5 with separations determined by $c_{4}$
\cite{Aldegunde:2008}. For excited rotational states, however, only $F$ is
conserved and $I$ is a poor quantum number.

An external magnetic field splits each rotational manifold into
$(2N+1)(2I_{\text{Rb}}+1)(2I_{\text{Cs}}+1)$ Zeeman-hyperfine sublevels, so
there are 32 levels for $N=0$ and 96 levels for $N=1$. Assignment of quantum
numbers to the individual hyperfine levels is non-trivial and depends on the
magnetic field regime \cite{Aldegunde:2008}. The field mixes states with
different values of $F$ that share the same total projection $M_{F}$. At low
field, the levels are still approximately described by $F$ and $M_F$
(equivalent to $I$ and $M_I$ for $N=0$). At high field, however, the nuclear
spins decouple and the individual projections $M_{N}$, $m^{\text{Rb}}_{I}$ and
$m^{\text{Cs}}_{I}$ become nearly good quantum numbers, with
$M_{F}=M_{N}+m^{\text{Rb}}_{I}+m^{\text{Cs}}_{I}$.
%The only quantum number that is good in all field regimes is $M_{F}$.

\begin{figure}
\includegraphics[width=0.49\textwidth]{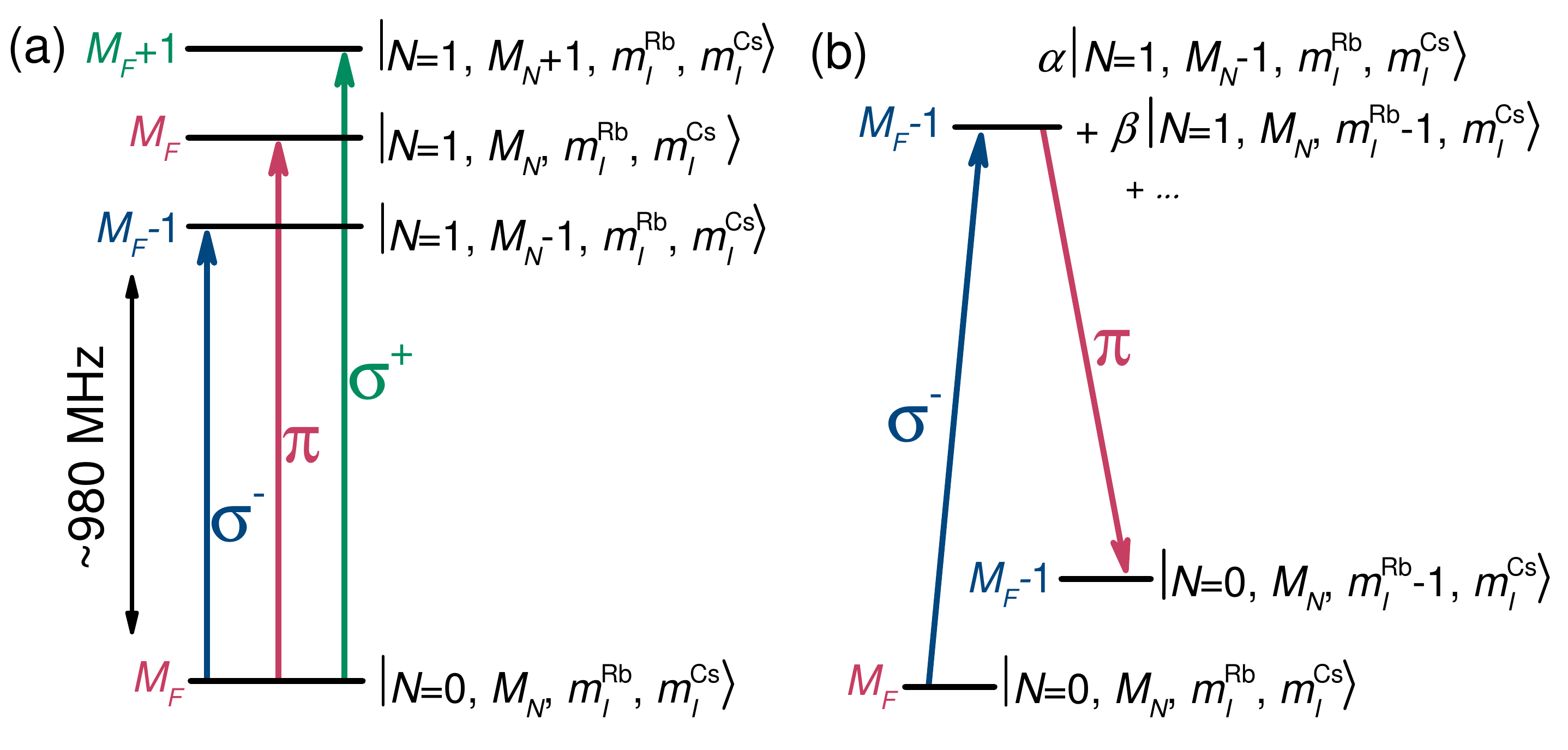}
\caption{\label{fig:SelectionRules} (Color online) Electric dipole transitions
in $^{87}$Rb$^{133}$Cs between the $N=0$ and $N=1$ rotational levels in the
vibrational ground state. (a) First-order allowed electric-dipole transitions
keep the hyperfine state projections from the atomic nuclear angular momentum
unchanged ($\Delta m^{\text{Rb}}_{I} = \Delta m^{\text{Cs}}_{I} = 0$). In the
absence of hyperfine interactions we would therefore expect to be able to drive
3 transitions with $\Delta M_{F} = \Delta M_{N} = 0, \pm1$ for microwave
polarizations $\pi, \sigma^{\pm}$. (b) Scheme for changing nuclear angular
momentum in the rotational ground state. Interactions involving the nuclear
electric quadrupole moments of $^{87}$Rb and $^{133}$Cs mix quantum states with
different nuclear spin quantum numbers $m^{\text{Rb}}_{I}$ and
$m^{\text{Cs}}_{I}$. We use a two-photon pulse sequence to transfer up to a
mixed excited quantum state in which both the initial and desired
values of $M_N$ are present.}
\end{figure}

%Discussion of selection rules
A microwave field induces electric dipole transitions between rotational
levels. At low field, all transitions allowed by the selection rules $\Delta
F=0,\pm1$ and $\Delta M_F=0,\pm1$ have significant intensity. At higher field,
however, additional selection rules emerge. If hyperfine couplings are
neglected, electric dipole transitions leave the nuclear spin states unchanged
($\Delta m^{\text{Rb}}_{I}=\Delta m^{\text{Cs}}_{I}=0$) and are allowed only
between neighboring rotational states such that $\Delta N=\pm1$, $\Delta
M_{N}=0, \pm1$ for microwave polarizations $\pi, \sigma^{\pm}$. In the absence
of hyperfine interactions (where $M_{N}$ would be a good quantum number) we
would be able to drive at most three transitions from any given hyperfine
level, as shown in Fig.~\ref{fig:SelectionRules}(a). Nuclear
quadrupole coupling in the $N=1$ state mixes levels with different values of
$M_{N}$, $m^{\text{Rb}}_{I}$ and $m^{\text{Cs}}_{I}$, and additional
transitions become allowed. The relative strengths of the transitions depend
on the magnitude of the component of the destination state that preserves
$m^{\text{Rb}}_{I}$ and $m^{\text{Cs}}_{I}$~\cite{Aldegunde:2009}. The presence
of state mixing allows us to use a multi-photon scheme to move the population
to different hyperfine states of the rotational ground state.
Fig.~\ref{fig:SelectionRules}(b) shows an example of the simplest possible
variation of the scheme using two microwave photons to change the hyperfine
state by $\Delta M_{F}=-1$.

\begin{figure*}
\includegraphics[width=\textwidth, trim={1.2cm, 1.2cm, 1.2cm, 1.2cm}]{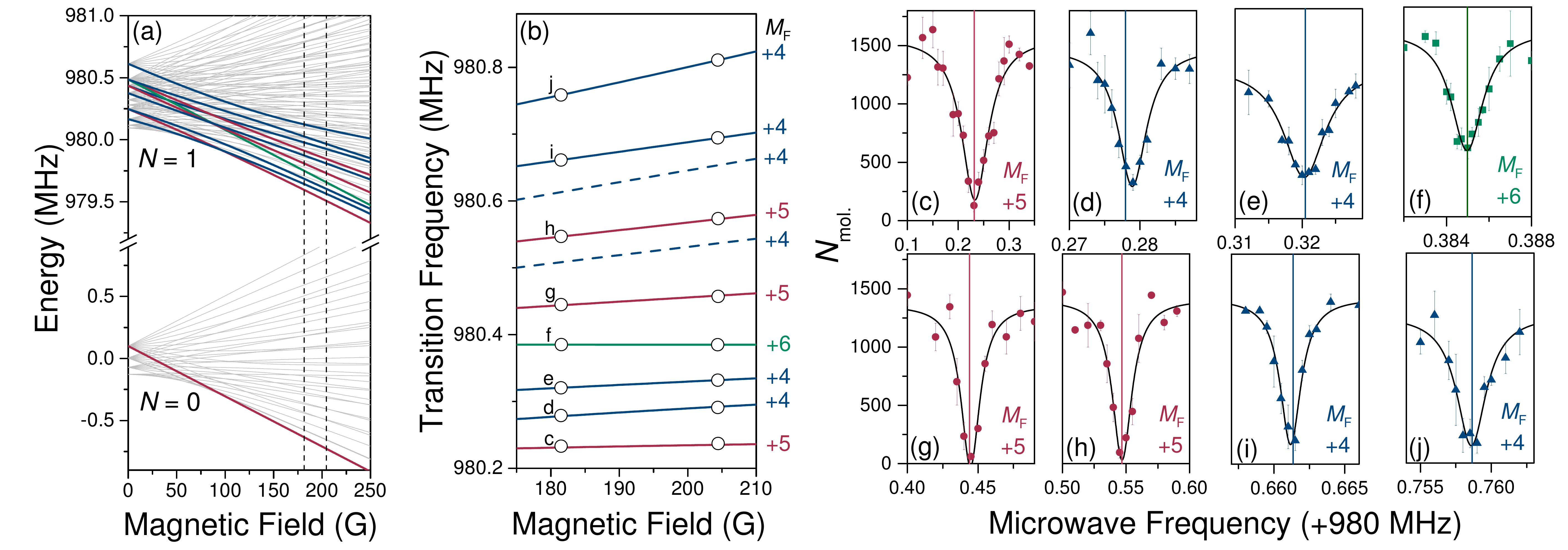}
\caption{\label{fig:Spectroscopy} (Color online) Spectroscopy of the
first-excited rotational state. (a) Hyperfine Zeeman structure of the $N = 0$
and $N = 1$ states. The initial state in $N = 0$ is highlighted as the bold red
line. The 10 states that are accessible from this initial state in $N = 1$ are
shown as bold blue ($M_{F}=4$), red ($M_{F}=5$), and green ($M_{F}=6$) lines.
The vertical dotted lines mark the two magnetic fields at which spectroscopy is
performed in this work. (b) Comparison of experimentally measured transition
frequencies from $\ket{N=0, M_{F}=5}$ to $\ket{N=1, M_{F}=4,5,6}$ with the
fitted theory. Dashed lines indicate transitions that are weakly allowed but we
have not observed. Error bars are not visible at this scale (see
Table~\ref{table:Spectroscopy}). (c-j) Spectra of all the transitions found in
this work at a magnetic field of $\sim181.5$~G. The vertical lines shows the
position of each transition given by fit to the results. The widths of all of
the features are Fourier-transform limited. The pulse duration used is less
than a $\pi$ pulse for each transition. The specific pulse durations are (c)
\SI{12}{\micro\second}, (d) \SI{150}{\micro\second}, (e)
\SI{100}{\micro\second}, (f) \SI{400}{\micro\second}, (g)
\SI{60}{\micro\second}, (h) \SI{50}{\micro\second}, (i)
\SI{400}{\micro\second}, (j) \SI{200}{\micro\second}.}
\end{figure*}

%Brief experimental outline
Our experimental apparatus and method for creating ultracold
$^{87}$Rb$^{133}$Cs molecules have been discussed in previous
publications~\cite{Harris:2008, Jenkin:2011, Cho:2011, McCarron:2011,
Koppinger:2014, Molony:2014, Gregory:2015}; we will therefore give only a brief
overview here. We begin by using magnetoassociation on a magnetic Feshbach
resonance to create weakly bound molecules from an ultracold atomic mixture
confined in a crossed-beam optical trap
($\lambda=1550$~nm)~\cite{Koppinger:2014}. We remove the remaining atoms by
means of the Stern-Gerlach effect, leaving a pure sample of trapped molecules.
These molecules are then transferred to a single hyperfine state of the
rovibational ground-state by stimulated Raman adiabatic passage
(STIRAP)~\cite{Molony:2014, Molony:2016b}. In this work, we create a sample of
up to $\sim$2000 $^{87}$Rb$^{133}$Cs molecules in the lowest hyperfine state
(shown in Fig.~\ref{fig:Spectroscopy}(a)) at a temperature of
\SI{1.17(1)}{\micro\kelvin} and peak density of
$8.1(8)\times10^{10}$~cm$^{-3}$. It is important to note that in order to
measure the number of molecules in our experiment we reverse both the STIRAP
and magnetoassociation steps and subsequently use absorption imaging to detect
the atoms that result from the molecular dissociation. Throughout, therefore,
we always measure the number of molecules in the hyperfine state initially
populated by STIRAP.

%Specific apparatus for spectroscopy
Our apparatus is equipped with two omnidirectional $\lambda/4$ antennas placed
close to the outside of the fused silica cell. The polarization from each is
roughly linear at the position of the molecules. They are oriented
perpendicular to each other and aligned with respect to the direction of the
static magnetic field such that one preferentially drives transitions with
$\Delta M_{F}=0$ and the other drives those with $\Delta M_{F}=\pm1$. Each
antenna is connected to a separate signal generator, which is frequency
referenced to an external 10~MHz GPS reference. Fast ($\sim$ns) switches are
used to generate microwave pulses of well-defined duration (typically
\SI{1}{\micro\second}~-~\SI{500}{\micro\second}).

\begin{table}
\vspace{0.5cm}
\begin{tabular*}{\linewidth}{@{\extracolsep{\fill}}llllr}
\hline
\hline
$M_{F}$ & $B$~(G) & $f_{\text{The.}}$~(kHz) & $f_{\text{Exp.}}$~(kHz) & $\Delta f$~(kHz)\\
\hline
\hline
+5 & 181.507(2) & 980~231.07 & 980~233(2) & $-$2(2) \\
~  & 204.436(2) & 980~235.14 & 980~237(1) & $-$2(1) \\
\hline
+4 & 181.484(1) & 980~277.96 & 980~278.9(2) & $-$0.9(2) \\
~  & 204.397(2) & 980~292.08 & 980~291.0(2) & 1.1(2) \\
\hline
+4 & 181.487(1) & 980~320.47 & 980~320.4(2) & 0.1(2) \\
~  & 204.397(2) & 980~331.83 & 980~331.8(3) & 0.0(3) \\
\hline
+6 & 181.541(2) & 980~384.98 & 980~384.97(6) & 0.01(6) \\
~  & 204.38(1) & 980~384.87 & 980~384.90(5) & $-$0.03(5) \\
\hline
+5 & 181.507(2) & 980~443.97 & 980~444.8(7) & $-$0.8(7) \\
~  & 204.436(2) & 980~458.35 & 980~457.2(8) & 1.1(3) \\
\hline
+5 & 181.507(2) & 980~546.75 & 980~546.9(7) & $-$0.2(7) \\
~  & 204.436(2) & 980~572.86 & 980~573.5(6) & $-$0.6(6) \\
\hline
+4 & 181.487(1) & 980~661.35 & 980~661.15(6) & 0.20(6) \\
~  & 204.397(2) & 980~694.22 & 980~694.35(5) & $-$0.13(5) \\
\hline
+4 & 181.487(1) & 980~758.64 & 980~758.6(1) & 0.0(1) \\
~  & 204.397(2) & 980~810.62 & 980~810.8(3) & $-$0.2(3) \\
\hline
\hline
\end{tabular*}
\caption{\label{table:Spectroscopy}Microwave transitions found in
$^{87}$Rb$^{133}$Cs from $\ket{v=0,N=0}$ to $\ket{v=0, N=1}$. All transitions
start from the spin-stretched $M_{F}=+5$ hyperfine level of the rotational
ground state. Each transition is labeled by the $M_{F}$ quantum number of the
destination hyperfine level in the first-excited rotational state.}
\end{table}

%Spectroscopy method
The large dipole moment of the molecule (1.225~D~\cite{Molony:2014}) makes it
easy to drive fast Rabi oscillations between neighboring rotational states. To
perform the spectroscopy, therefore, we pulse on the microwave field for a time
($t_{\text{pulse}}$) which is less than the duration of a $\pi$-pulse for the
relevant transition ($<t_{\pi}$). We then observe the transition as an apparent
loss of molecules as they are transferred into the first-excited rotational
state. To avoid ac Stark shifts of the transition centers, the optical trap is
switched off throughout the spectroscopy; the transition frequencies are thus
measured in free space. We find that the widths of all of the features we
measure are Fourier-transform limited, i.e. the width is proportional to
$1/t_{\text{pulse}}$. We therefore iteratively reduce the power to get slower
Rabi oscillations and allow longer pulse durations. We also note that radically
different $t_{\text{pulse}}$ are required for each transition depending on the
transition strength and antenna used. We carry out the spectroscopy at two
different magnetic fields $\sim23$~G apart; the magnetic field is calibrated
using the microwave transition frequency between the $\ket{f=3,m_{f}=+3}$ and
$\ket{f=4,m_{f}=+4}$ states of Cs.

%Introduction of spectroscopy data
With the population initially in the lowest hyperfine level ($M_{F}=5$) of the
rovibrational ground state, we expect to find a maximum of 10 transitions to
the first-excited rotational state $\ket{N=1, M_{F}=4,5,6}$. We are able to
observe 8 of these transitions, at the frequencies given in
Table~\ref{table:Spectroscopy}. A complete set of spectra at a magnetic field
of $\sim181.5$~G is also shown in Fig.~\ref{fig:Spectroscopy}(c-j).
Calculations of the expected intensities of the two unseen transitions show
that the relative transition probability is $\sim10^{-4}$ lower than for those
we do observe.

\begin{table}
%\vspace{0.5cm}
\begin{tabular*}{\linewidth}{@{\extracolsep{\fill}}ccc}
\hline
\hline
Constant & Value & Ref. \\
\hline
\hline
$B_{v}$ &		490.155(5)~MHz & \cite{Fellows:1999}\\
~		&		490.173~994(45)~MHz & This Work\\
\hline
$D_{v}$	&	213.0(3)~Hz & \cite{Fellows:1999} \\
\hline
$(eQq)_{\text{Rb}}$	&  $-$872~kHz & \cite{Aldegunde:2008} \\
~ &	$-$809.29(1.13)~kHz & This Work\\
\hline
$(eQq)_{\text{Cs}}$	&	51~kHz & \cite{Aldegunde:2008}\\
~	&	59.98(1.86)~kHz & This Work\\
\hline
$c_{\text{Rb}}$ & 29.4~Hz & \cite{Aldegunde:2008} \\
\hline
$c_{\text{Cs}}$ & 196.8~Hz & \cite{Aldegunde:2008} \\
\hline
$c_{3}$ & 192.4 Hz & \cite{Aldegunde:2008} \\
\hline
$c_{4}$	&	17.3~kHz & \cite{Aldegunde:2008} \\
~	&	19.019(105)~kHz & This Work\\
\hline
$g_{r}$ & 0.0062  &  \cite{Aldegunde:2008} \\
%\hline
%$\sigma_{\text{Rb}}$ & 3531~ppm & \cite{Aldegunde:2008} \\
%\hline
%$\sigma_{\text{Cs}}$ & 6367~ppm & \cite{Aldegunde:2008} \\
\hline
$g_{\text{Rb}} \cdot (1-\sigma_{\text{Rb}})$	&	1.8295(24) & This Work\\
\hline
$g_{\text{Cs}} \cdot (1-\sigma_{\text{Cs}})$	&	0.7331(12) & This Work\\
\hline
\hline
\end{tabular*}
\caption{\label{table:Fitting} Constants involved in the molecular Hamiltonian
for $^{87}$Rb$^{133}$Cs. Parameters not varied in the least-squares fit are
taken from the literature. The majority of the fixed terms are calculated using
density-functional theory (DFT)~\cite{Aldegunde:2008}, with the exception of
the centrifugal distortion constant $D_v$, which is obtained from laser-induced
fluorescence combined with Fourier transform spectroscopy
(LIF-FTS)~\cite{Fellows:1999}).}
\end{table}

%Fitting of the spectroscopy data
We fit our model to the experimental spectra by minimizing the sum of the
squared quotients between each residual and the uncertainty of the line. We fit
the rotational constant, nuclear quadrupole constants and scalar nuclear
spin-spin constant. The nuclear $g$-factors and shielding coefficients are
multiplied together in the Hamiltonian so it is not possible to separate them,
and we therefore fit the shielded $g$-factors
$g_{\text{Rb}}\cdot(1-\sigma_{\text{Rb}})$ and
$g_{\text{Cs}}\cdot(1-\sigma_{\text{Cs}})$. The resulting values, along with
the values of parameters held fixed at theoretical values, are given in
Table~\ref{table:Fitting}.

The fitted hyperfine parameters in Table~\ref{table:Fitting} are all within
10\% of the values predicted from DFT calculations~\cite{Aldegunde:2008},
except for $(eQq)_{\text{Cs}}$, which is about 15\% larger than calculated.
This helps to calibrate the probable accuracy of the calculations for other
alkali-metal dimers. The fitted value $c_4=19.0(1)$~kHz removes one of the two
largest sources of error in our recent determination of the binding energy
$D_0$ of $^{87}$Rb$^{133}$Cs in its rovibrational ground
state~\cite{Molony:2016}; the zero-field hyperfine energy of the $M_F=5$ state
is (21/4)$c_4$, which increases from 90(30)~kHz in ref.\ \cite{Molony:2016} to
99.9(6)~kHz. This increases the binding energy of the hyperfine-weighted
vibronic bound state by 9~kHz, giving a revised value $D_0= h \times
114\,268\,135.25(3)$~MHz. The fitted values of the shielded $g$-factors
$g_{\text{Rb}}\cdot(1-\sigma_{\text{Rb}})=1.829(2)$ and
$g_{\text{Cs}}\cdot(1-\sigma_{\text{Cs}})=0.733(1)$ are consistent with the
corresponding atomic values, $1.827\,232(2)$ \cite{White:1968} and
$0.732\,357(1)$ \cite{White:1973} (with the sign convention of Eqn.
\ref{eqn:Zeeman}). The latter include shielding due to the electrons in the
free atoms. Our values may be used in conjunction with the calculated molecular
shielding factors ($\sigma_{\text{Rb}}=3531$~ppm and
$\sigma_{\text{Cs}}=6367$~ppm \cite{Aldegunde:2008}) to obtain values of the
``bare" nuclear $g$-factors $1.836(3)$ and $0.738(1)$.

\begin{figure}
\includegraphics[width=0.5\textwidth]{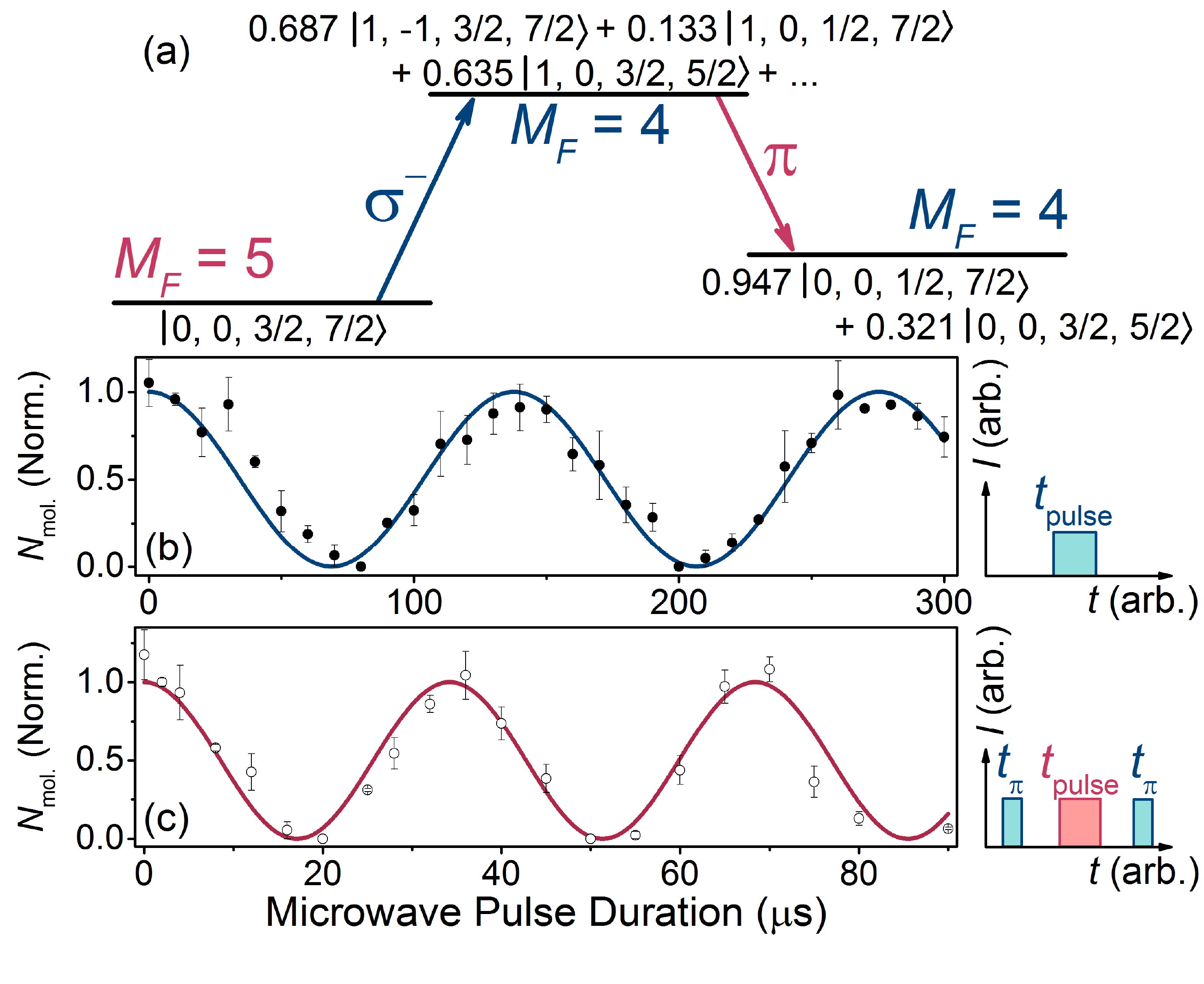}
\vspace{-0.95cm}

\caption{\label{fig:TwoPhotonTransfer} (Color online) Coherent population
transfer of molecules between hyperfine states in rotational states $N = 0$ and
$N = 1$. (a) Transfer scheme followed in this work. All molecules start in the lowest hyperfine level ($M_{F}=5$) of the
rovibrational ground state. State notation is given by $\ket{N, M_{N},
m_{I}^{\text{Rb}}, m_{I}^{\text{Cs}}}$. (b) One-photon transfer of molecules to
a single hyperfine level of the $N = 1$ excited rotational state. Microwaves
with $\sigma^{-}$ polarization drive transitions to $\ket{N=1, M_{F}=+4}$. By
varying the duration of the microwave pulse, we observe Rabi oscillations to
determine the duration of a $\pi$-pulse allowing complete population transfer
to the excited rotational state. (c) Two-photon transfer to change the
hyperfine level populated in the rovibrational ground state. Once the
population has been transferred to the $N=1$ excited state, we introduce a
second microwave pulse with different frequency and polarization from the
first. This drives the population back down to a different hyperfine level of
the rovibrational ground state than that initially populated.}
\end{figure}

%Population transfer - Scheme
The STIRAP transfer produces molecules in a spin-stretched state, where $\lvert
m^{\text{Rb}}_{I}+m^{\text{Cs}}_{I} \rvert$ has its maximum possible value and
$M_{N}, m^{\text{Rb}}_{I}, m^{\text{Cs}}_{I}$ are all good quantum numbers.
However, the other hyperfine states of both $N=0$ and 1 are significantly mixed
in the uncoupled basis set at the fields considered here, and have no good
quantum numbers other than $M_{F}$. In Fig.~\ref{fig:TwoPhotonTransfer}, we
demonstrate complete transfer of the molecular population between these
mixed-character hyperfine states. We begin by transferring the molecules to an
$M_{F}=4$ level of $N=1$ (transition frequency = $980 320.47$~kHz, shown in
Fig.~\ref{fig:Spectroscopy}(e)). The eigenvector component of the uncoupled
basis function that couples to our initial $N=0$ hyperfine level is
$\sim0.687$. With the microwave power available, $\pi$-pulses on this
transition can be driven with pulse durations \SI{<10}{\micro\second}, though
it is important when using short pulses that the separation between available
states is greater than the Fourier width of the pulse. We reduce the microwave
power such that the Rabi frequency of the transition is $2\pi \times
7.26(5)$~kHz, as shown in Fig.~\ref{fig:TwoPhotonTransfer}(b), ensuring that we
do not couple to neighboring transitions. Single $\pi$-pulses allow complete
transfer of the population to the destination hyperfine level. We subsequently
transfer the molecules to a different hyperfine level of $N=0$ by applying a
second microwave field with a different polarization and frequency. We choose
to use $\pi$-polarized microwaves to transfer the molecules to the
higher-energy of the two $M_{F}=4$ levels of $N=0$ (transition frequency =
$980119.14$~kHz). At this field, the composition of this final level is
$0.947\ket{4,0,1/2,7/2}+0.321\ket{4,0,3/4,5/2}$ in the uncoupled basis $\ket{N,
M_{N}, m^{\text{Rb}}_{I}, m^{\text{Cs}}_{I}}$. We observe Rabi oscillations on
the second transition by pulsing on the $\pi$-polarized microwaves in between
two $\pi$-pulses on the $\sigma^{-}$-polarized microwave transition as shown in
Fig.~\ref{fig:TwoPhotonTransfer}(c). Coherent transfer is achieved with a Rabi
frequency of $2\pi \times 29.2(3)$~kHz.

%Conclusion/Summary
In summary, we have performed high-precision microwave spectroscopy of
ultracold $^{87}$Rb$^{133}$Cs molecules in the vibrational ground state, and
have accurately determined the hyperfine coupling constants for the molecule.
Our results confirm that the hyperfine coupling constants calculated by
Aldegunde \emph{et al.}\ \cite{Aldegunde:2008} are generally accurate to within
$\pm10$\%, calibrating the probable accuracy of the calculations for other
alkali-metal dimers. The resulting understanding of the hyperfine structure
enables full control of the quantum state, as illustrated by our demonstration
of coherent transfer to a chosen hyperfine state in either the first-excited or
ground rotational state. Such complete control is essential for many proposed
applications of ultracold polar molecules, and opens the door to a range of
exciting future experimental directions, including studies of quantum
magnetism~\cite{Barnett:2006, Gorshkov:2011} and novel many-body
phenomena~\cite{Carr:2009, Baranov:2012}.

\begin{acknowledgments}
This work was supported by the U.K. Engineering and Physical Sciences Research
Council (EPSRC) Grants No. EP/H003363/1, EP/I012044/1, and GR/S78339/01. JA
acknowledges funding by the Spanish Ministry of Science and Innovation Grants
No.\ CTQ2012-37404-C02, CTQ2015-65033-P, and Consolider Ingenio 2010
CSD2009-00038. The experimental results and analysis presented in this paper
are available at DOI:10.15128/r12j62s485j.
\end{acknowledgments}

\bibliography{RbCsReferences}

\end{document}